\documentclass[journal]{IEEEtran}
\usepackage{bm}
\usepackage{listings}
\usepackage{amsmath, verbatim,graphicx}
\usepackage{amsfonts, dsfont, bbold}
\usepackage{tikz}
\usetikzlibrary{shapes,arrows,positioning}
\usepackage{color}

\usepackage{todonotes}
\usepackage[skins,listings]{tcolorbox}
\newcommand{\ket}[1]{|#1\rangle}

\usepackage[normalem]{ulem} 

\begin{document}
\title{Hybrid Programming for Near-term Quantum Computing Systems}
\author{A.~J.~McCaskey,~E.~F.~Dumitrescu, D.~I.~Liakh,
        and~T.~S.~Humble\\Quantum Computing Institute, Oak Ridge National Laboratory\\Oak Ridge, Tennessee, 37831
\thanks{This manuscript has been authored by UT-Battelle, LLC, under contract DE-AC05-00OR22725 with the US Department of Energy (DOE). The US government retains and the publisher, by accepting the article for publication, acknowledges that the US government retains a nonexclusive, paid-up, irrevocable, worldwide license to publish or reproduce the published form of this manuscript, or allow others to do so, for US government purposes. DOE will provide public access to these results of federally sponsored research in accordance with the DOE Public Access Plan. (http://energy.gov/downloads/doe-public-access-plan).}}

\markboth{International Conference on Rebooting Computing, ~May~2018}%
{}
\maketitle
\begin{abstract}
Recent computations involving quantum processing units (QPUs) have demonstrated a series of challenges inherent to hybrid classical-quantum programming, compilation, execution, and verification and validation. Despite considerable progress, system-level noise, limited low-level instructions sets, remote access models, and an overall lack of portability and classical integration presents near-term programming challenges that must be overcome in order to enable reliable scientific quantum computing and support robust hardware benchmarking. In this work, we draw on our experience in programming QPUs to identify common concerns and challenges, and detail best practices for mitigating these challenge within the current hybrid classical-quantum computing paradigm. Following this discussion, we introduce the XACC quantum compilation and execution framework as a hardware and language agnostic solution that addresses many of these hybrid programming challenges. XACC supports extensible methodologies for managing a variety of programming, compilation, and execution concerns across the increasingly diverse set of QPUs. We use recent nuclear physics simulations to illustrate how the framework mitigates programming, compilation, and execution challenges and manages the complex  workflow present in QPU-enhanced scientific applications. Finally, we codify the resulting hybrid scientific computing workflow in order to identify key areas requiring future improvement.
\end{abstract}


\IEEEpeerreviewmaketitle

\section{Introduction}
Currently available quantum processing units (QPUs) consisting of tens of qubits are providing a unique capability for understanding hybrid classical-quantum algorithms and associated speedups for future scientific computing applications. Such applications range across scientific domains, and small-scale demonstrations of quantum programming have been developed in fields such as nuclear and high-energy physics, machine learning, chemistry and materials science \cite{ascrreport}. While QPU hardware development is progressing at a rapid pace, these near-term quantum computing systems are far from ideal \cite{ascrreport2017}. Low-level unitary quantum instruction noise, readout errors, decoherence patheways, and remote programming access models limit the scalability of these devices to research applications. Recent results in hybrid classical-quantum variational algorithms demonstrate the potential ability to mitigate some of these challenges, specifically QPU noise errors, but there is an overall lack of awareness of the software tooling needed by programmers and domain scientists to leverage such computing systems in a robust and coherent way. 
\par
State of the art demonstrations of hybrid scientific quantum computations on gate-model QPUs, e.g. devices offered by vendors such as IBM, Rigetti, and Google, have had varied success and limited simulation accuracy \cite{linke2017experimental,kandala2017,deuteronarxiv,o2016scalable}. For example, the first variational hybrid quantum computation via remote cloud resources reached an overall simulation accuracy of 3 percent \cite{deuteronarxiv}. This is a computation that can be performed by a classical computer in microseconds, yet it took months of work to map it to a quantum computer and execute via a remote cloud access model with job queue constraints. These types of near-term hybrid programming challenges must be overcome to enable reliable and reproducible quantum computing applications and to support the continued testing and characterization of quantum computer performance.
\par
Can a robust software platform and workflow improve the usability and accessibility of near-term quantum computers for scientific applications? In this work, we attempt to answer this question in the affirmative by providing a model workflow for near-term hybrid quantum computations that enables useful scientific applications and makes quantum computing technologies accessible to a broader community. We will detail the primary challenges present in these hybrid, noisy quantum computations and discuss best practices for addressing them. We describe the XACC quantum compilation and execution framework that provides a hardware and language agnostic solution for many of these challenges and supports extensible methodologies that provide a strategy for managing programming concerns across an increasingly diverse set of QPU tools. As an example, we discuss recent demonstrations of scientific applications built on XACC for nuclear physics.
\par
This work is organized as follows: first, we provide a discussion of hybrid classical-quantum computing systems, including high-level discussions of requisite programming and execution models. We then discuss the near-term challenges present in programming, compiling, and executing hybrid scientific computing applications. Afterwards we introduce the XACC quantum programming framework and detail how it addresses these unique challenges. Finally, we conclude by demonstrating its utility with the example of computing the binding energy of the deuteron bound state.
\section{Hybrid Computing Systems}
We broadly define hybrid computing systems as a class of abstract machine models that combine different computational paradigms. However, we specialize our analysis to the case of realizable architectures that integrate a conventional classical Turing machine with a quantum Turing machine. Research into these different machine models has clarified that they are not equivalent with respect to computational power \cite{simon1997power}, and we address some of the unique considerations that arise by using the conventional model to program and control the quantum model.
\begin{figure}[ht]
\centering
\includegraphics[width=2.5in]{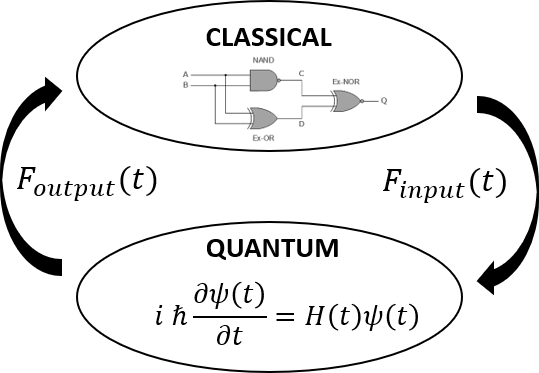}
\caption{A hybrid computing infrastructure that integrates classical systems of Boolean logic into the control of quantum dynamics requires an interface for input-output (IO) based on the transmission and reception of physical fields $F_{input}(t)$ and $F_{output}(t)$.}
\label{fig:hybrid}
\end{figure}
\par
As shown in Fig.~\ref{fig:hybrid}, the two-way interaction between a classical and a quantum machine model imposes a constraint that requires the machines to share a common understanding of language. In practice, this language is expressed as the physical fields that define the operations issued by the conventional model and implemented within the quantum domain \cite{BrittICRC2017}. Electric, magnetic, and optical fields are prominent examples by which the current technology controls the quantum physical systems. It is notable that this description is inherently analog due to the continuity of the time-dependent Schrodinger equation. Translation between the system of Boolean logic characterizing the conventional Turing machine into the analog fields is necessarily limited by the available computational power. In practice, this amounts to constraints imposed by the available digital-to-analog converters and the range of the arbitrary waveform generators as well as the speed at which the logical network can be processed and the connectivity of the control system.
\par
The outstanding concern for programming such a hybrid computing system is the controllability of the conventional and quantum machines \cite{brown2016co,corrigan2017quantum}. That is to say, for those scenarios in which the quantum machine must execute a series of issued instructions, how accurately are these instructions realized and how precisely does the result reflects the effect of the intended instructions?
\subsection{Client-Server Model for Hybrid Computing Systems}
\label{sec:client}
Current state-of-the-art hybrid computing systems integrate existing CPU-based clients with QPU-based servers \cite{Devitt2016,britt2017high}. The latter represent the online availability of a programmable infrastructure to access the field generators that drive the control of a quantum physical device. The applied fields are shaped and scheduled to control the dynamics of an addressable array of quantum physical subsystems, which for convenience we denote generically as the quantum register. Similarly, the response fields emitted by the register elements are collected and discriminated to generate binary representations that characterize the state of the register.
\par
\begin{figure}[ht]
\centering
\includegraphics[width=2.5in]{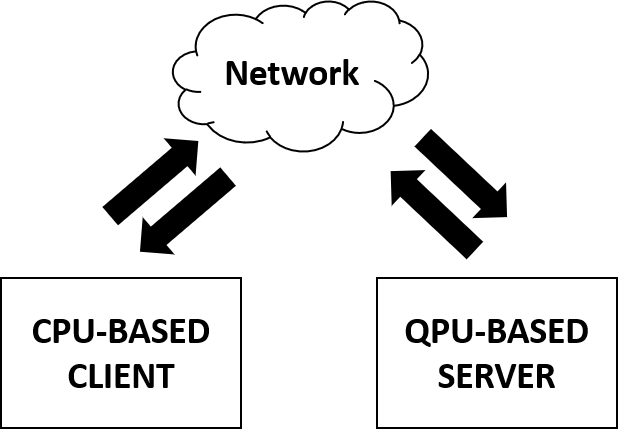}
\caption{Currently, most hybrid computing systems rely on a CPU-based client interacting with a QPU-based server over a network. The QPU-based server presents a classical control interface to the client that can be customized to a variety of programming styles. Internally, the QPU-based server must parse these instructions into a local representation that carries out the requested sequence of control fields.}
\label{fig:cpuqpumodel}
\end{figure}

A QPU-based server often includes conventional CPUs for purposes of parsing the digital programming instructions that generate the shape and timing of the control fields as well as the detection and discrimination of the response fields. Access to the QPU-based server requires an interface that may adhere to conventional logic, for example, as found in modern networking communication technology, to accept instructions from and return results to a CPU-based client. Currently, the conventional client-server model shown in Fig.~\ref{fig:cpuqpumodel} dominates the access method to QPU systems due in large part to the experimental nature of these machines.
\par
Within the client-server model for hybrid computing systems, the QPU represents a layer of language parsers that translate the Boolean logic of the client to the control fields required to drive the quantum register. As detailed elsewhere \cite{BrittICRC2017}, the QPU is partitioned into a control unit, execution units, and the register itself. The control unit parses the client instructions received by the server into the local instruction set architecture for generating and applying control fields. Application of the fields are carried out by the execution units, which in practice represent waveform generators for electric, magnetic and optical fields. A similar signal flow occurs for detecting output fields and generating a digital response.
\subsection{Programming and Execution Model}
The client-side API for the hybrid computing system dictates how users access the quantum physical devices, while the functionality of the QPU-based server is restricted to the local interpretation of these transmitted instructions. In this programming model, a client may select which instructions to send from a predefined instruction set for the QPU. The instruction set architecture (ISA) defines the functionality of the QPU and the language for the control unit \cite{Britt2017ISA}. These instructions are generally transmitted as character strings that are mapped by the QPU-based server into pre-compiled functions that execute the instruction. Calls to these libraries initiate the cascade of logic required to trigger the execution units, which subsequently apply the necessary fields to the register elements.
\begin{figure}[ht]
\centering
\includegraphics[width=1.0in]{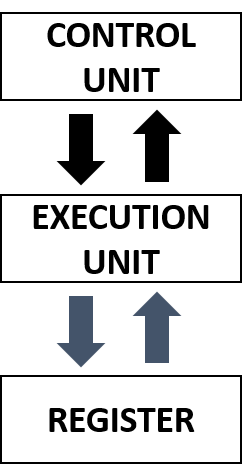}
\caption{Components in a QPU-based server include a control unit that expresses the instruction set architecture (ISA), multiple execution units that translate instructions into applied fields, and the quantum register which stores the computational state. This layered, hierarchical structure provides a natural separation of concerns but introduces challenges to programming near-term, noisy QPUs.}
\label{fig:execmodel}
\end{figure}
\par
As shown in Fig.~\ref{fig:execmodel}, the two-way information flow within the QPU-based server provides a natural separation of concerns for device programming. The logic dictating computational functionality is isolated within the control unit and determined by programs expressed within the ISA. Similarly, the execution unit hosts the translation of individual instructions into the fields that carry out the intended logical transformations on the register. The information unique to the implementation of either layer is notionally not required in the opposing layer. However, as we explore in the subsequent section, this design of the current QPU-based servers leads to several challenges for application programming.
\section{Programming Challenges for Noisy QPUs}
Despite the natural separation of concerns that arises in a hybrid computing system using a QPU-based server, this design presents a number of challenges when programming currently available noisy quantum processors. Advances in our understanding and engineering of QPUs have enabled demonstrations with register capacities up to 20 addressable elements, yet these realizations still demonstrate a non-trivial amount of noise relative to the underlying decoherence rates, greatly limiting the depth of quantum circuits with even modest reliability. This raises a need for repeated sampling of the circuit execution, but existing client-server access models hamper this interaction and greatly limit the overall utility and accessibility of the QPUs.
\par
Code portability is also a growing concern as many QPU vendors trend toward standalone programming frameworks that interact with their tightly controlled and proprietary ISA. These barriers to portability impede efforts to perform verification of software and benchmarking of hardware as well as methods to validate hardware behavior using numerical simulation. Finally, the individual workflow steps required to program, compile, and execute hybrid applications with near-term QPUs are tightly coupled. The lack of inter-operability across available software and hardware platforms raises a concern that the users programming hybrid computing systems will face artificial barriers to adoption and performance.
\par
We provide a more detailed description of each of these challenges with an analysis for how hybrid computation is affected. While several references provide background for additional technical details, we identity how each of these challenges must be overcome in order to advance utilization of noisy QPU systems.
\subsection{Gate Noise and Execution Errors}
\label{sec:noise}
In a QPU, gate noise represents a lack of control over the fundamental physical processes by which program instructions are executed. Among many measures of instruction accuracy, the quantum state fidelity quantifies the accuracy by which an observed state meets the instruction design requirements. For this characterization, a noisy gate applied to a well-defined quantum register state will induce a logical transform that prepares the register in a state that is not parallel with the expected outcome. Assuming only pure states, the fidelity is defined as the squared magnitude of the inner product between the observed and expected register states and, for noisy gates, it is always less then unity. A similar fidelity measure can be defined for mixed states in terms of the trace distance. 
\par
Noisy gate operation obviously influences program execution by diverting the computational state away from the intended algorithmic design. Quantum error correction and fault-tolerant gate protocols may be expected to mitigate this noise eventually, but such techniques are beyond the scope of current hardware devices. Instead, programming noisy QPUs must directly address the presence of errors within the application logic. This may be as simple as repeated execution of the program as a method of sampling the computer outputs, or it may be a more sophisticated redesign of the compiled circuit to mitigate against known errors. For such methods, the burden of understanding the noise that arises from the execution of these instructions lies on the programmer. However, as noted in the previous section, the separation of concerns for current QPU-based servers isolates the physics of the execution unit and register from the user. The ISA alone provides the interface for programming and controlling the computational logic.
\par
So far the application developers have relied on external characterization of the gate noise outside of the programming workflow. This requires consideration of errors during the algorithmic design stage which is largely based on manual analysis \cite{Temme2017}. Such strategies are untenable as circuit complexity increases. Programming models that are device-aware are currently lacking, but would be necessary to automate circuit rewriting techniques to compensate for noisy gates.
\par
In addition to the effects of gate noise, execution errors for a quantum program also arise from state-preparation and measurement (SPAM) errors. These errors correspond to faulty initialization or measurement of a register element. For example, an instruction to initialize the computational state $\ket{0}$ may inadvertently prepare the state $\ket{1}$ or a superposition of these two possibilities. Similarly, measurement of the state $\ket{0}$ may instead project into $\ket{1}$, which would be interpreted as the 1. It is notable that SPAM errors currently dominate QPU performance in multiple technologies with error rates nearly an order of magnitude above typical gate errors.
\par
Several strategies exist for mitigating SPAM errors such as using a series of repeated program executions to decode the correct result based on the maximum likelihood statistics. However, this example of statistical detection based on estimates of the output measurement values not only adds to the complexity of the program execution but also requires accurate characterization of the error mechanisms. Because of the relatively high SPAM error rates, a large number of samples are typically required to gain high confidence in the program behavior. But the separation of concerns between the physical and logical layers in the QPU details hides these physical errors from the programmer. Inline calibrations for initialization and measurement gates may be able to bridge this gap when the number of register elements is very small and the errors are independent, but cross-talk during measurement may invalidate the latter model. Moreover, such calibrations are intractable as the size of the register reaches ever larger sizes.
\subsection{Access Models}
\label{sec:access}
The overall system infrastructure required to operate current QPU systems is based largely on sensitive, experimental devices that cannot be easily distributed. Many vendors and laboratories therefore enable users to access these complex computing systems via remote, online server. Examples of these remote access models include QPU-based servers that support REST APIs that delegate request to a job queue service, which then schedules program execution on the QPU. In addition, many systems provide web portals that permit manual input methods for programming. These may also be operated via the REST API or Pythonic frameworks by creating batch-style program executions.
\par
This remote user access infrastructure suffices for small-scale program executions, but production-level computing, including scientific computing applications based on quantum acceleration, are not amenable to remote access models. As described in the subsequent sections, many current demonstrations of hybrid variational quantum algorithms employ repeated program executions \cite{Peruzzo2014,McClean2016theory,Li2017,Shen2017}. These methods require many consecutive serial executions in which each execution influences the next iteration of the algorithm. A remote network connection is therefore an impractical access model for this type of iterative hybrid algorithms because of the additional overhead associated with the remote calls, the communication parsing, and the queuing latency.
\par
A related challenge for QPU-based server programming is that the job queue service employed by most servers is based on individual serial QPU executions. This introduces a hardware bottleneck for the variational sampling algorithms that use multiple circuit executions to estimate a single observable. Related circuit executions are not collated within the queue, which slows down the overall application execution time. In contrast, job schedulers for typical high-performance computing systems operate by queuing a complete application execution. That is to say, when a job reaches the top of the queue, the associated applications are executed completely to remove unnecessary uncertainty in both the application and machine state.

\subsection{Portability}
\label{sec:portability}
A persistent concern for any application developer is the portability of existing code onto new platforms. Despite the major conceptual shift offered by the quantum computational model, this concern is also faced by the nascent hybrid application developer community. Many QPU vendors support standalone programming frameworks that interact only with their tightly controlled and proprietary ISA.  Although tight control over the system may offer advantages to the vendor, a major disadvantage to users is a need to retool for each QPU. Efforts to retool often slow down overall productivity and may, eventually, impede efforts to adopt new software or hardware.
\par
The lack of portability also presents a challenge for verification of software and benchmarking of hardware as well as for validation of the hardware behavior based on numerical simulations. In particular, there is a concern that test cases and benchmarks devised for one hardware platform will be incompatible with similar efforts developed for other platforms. In essence, developers cannot begin to address quantum benchmarking concerns without some form or mechanism of quantum code portability. Code verification and program validation represent important steps in certifying an application as correct. The above challenges for current quantum programming tools directly impact the ability for application developers to perform verification and validation. Noise and errors undermine the testability of the application, while non-portable codes challenge the ability to use different platforms for comparative analysis.

\subsection{Classical-Quantum Integration}
Primarily due to the remote access models employed and an overall lack of code portability, currently available quantum computing resources lack direct integration with conventional software and processing workflows. This integration is not necessary for experimental proof-of-concept demonstrations, in which the focus is on manual, device-level benchmarking and validation of the QPU physical behavior \cite{michielsen2017benchmarking}. But integration is necessary to make such devices available for development and testing of classical-quantum hybrid algorithms. Currently, these interactions are largely avoided or treated as separate stages of a manual workflow.
\par
There are several efforts underway to integrate quantum and classical workflows. Examples of domain-specific languages have appeared including Quipper \cite{Green2013}, ProjectQ \cite{Steiger2018projectqopensource}, Liquid \cite{Wecker2014}, and Q\# \cite{Svore2018qsharp} among many others. However, these languages target programmers with detailed knowledge and understanding of quantum logic, especially in the gate or circuit model. The integration of these languages with existing programming methods, including well-known languages such as C/C++ and python, are established by using externally reference library functions or embedding in a host language. In both cases, the understanding of quantum functionality and device behavior is ambiguous to the high-level language user and challenges the understanding of information flow and error analysis needed for profiling an application on a noisy QPU.


\subsection{Quantum Compilation Workflow}
\label{sec:workflow}
The quantum compilation workflow, like any compiler workflow, can be composed of discrete steps that at a black-box level read in a quantum source code and produce machine-level instructions ready for execution. However, along the way, many individual tasks need to be executed in order to ensure that the computation is amenable for the chosen hardware and is at least partly resilient to system noise and errors. This includes register allocation, instruction scheduling and layout, and, for noisy QPUs, pre- and post-processing of erroneous inputs and outputs.
\par
These individual compiler layers for quantum computation must be tightly coupled in that the layers closer to the high-level source code directly influence layers at a lower level closer to the hardware. Compilation layers can have a multiplicity greater than 1, meaning that multiple processes of the same type may be executed for a given layer, and these processes may or may not commute. Also, compilation layers may dictate whether or not any post-execution actions must take place.

\begin{figure}[ht]
\centering
\includegraphics[width=3.3in]{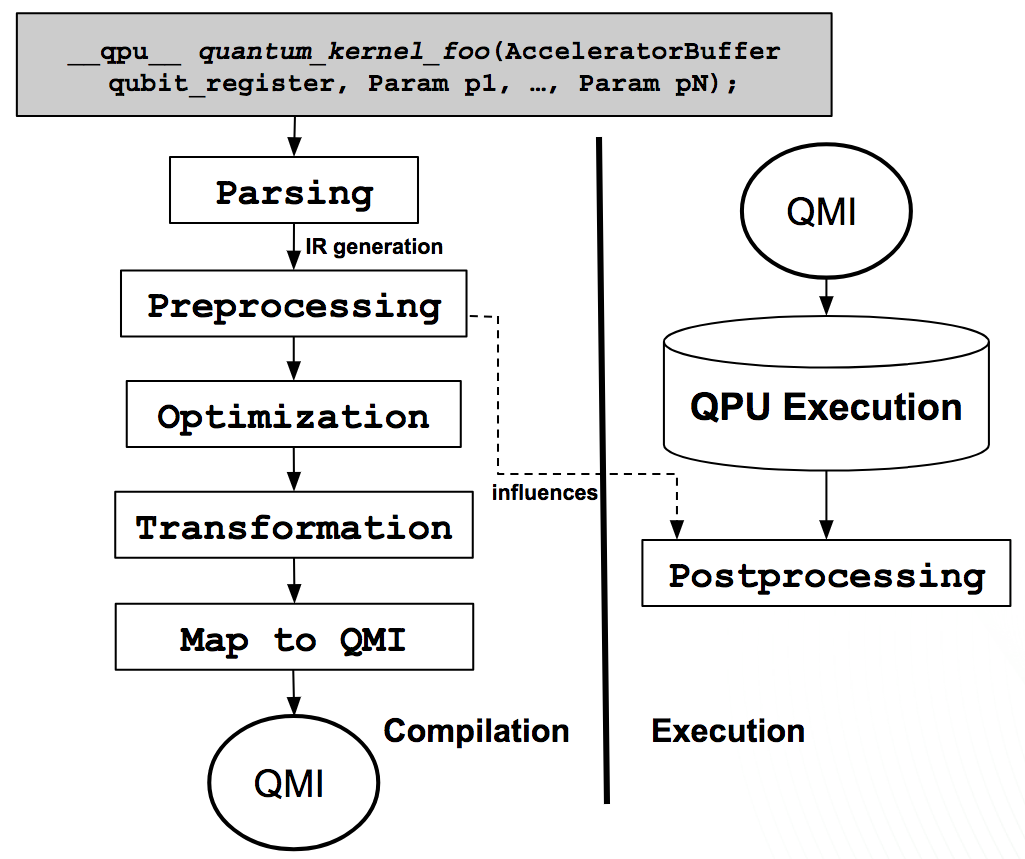}
\caption{The XACC compilation workflow is a series of processing layers amenable to near-term quantum computer programs. These layers include language parsing, source code pre-processing, optimization, and transformation. For near-term program behavior, the pre-processing layer is strongly tied to the post-processing methods that follow QPU execution, e.g., SPAM error mitigation strategies.}
\label{fig:workflow}
\end{figure}
\par
We provide working examples of the layers required the compilation workflow for noisy QPUs in Fig.~\ref{fig:workflow}. The first step in this workflow is source code parsing, which maps quantum code to an intermediate representation (IR) of the input program. This representation then passes through a compilation layer that provides generic pre-processing of the intermediate representation. The subsequent two layers target IR optimization and code transformation. Optimization may include simplification and enhancements to the programmed quantum logic which is isomorphic with the original program, while transformations modify the intermediate representation structure to satisfy logical constraints exposed by the hardware topology. The next layer maps the IR to the low-level quantum machine instructions (QMI) for the target hardware. The resulting representation is an executable that will be executed on the QPU-based server. Following execution, additional post-processing steps may be included to take advantage of the pre-processing layer. 
\par
These various layers of the compilation workflow must be tightly coupled, ideally through the use of a common IR to minimize overhead. Pre-processing directly influences the optimizations and transformations that may be implemented as well as the post-analysis of the QPU measurement results. At each compilation layer, multiple invocations of individual, yet different, pre-processors, optimizers, or transformations, and the execution of these processes may not commute with one another. For example, one preprocessor may update the IR instance in a way that makes a future pre-processor execution invalid. Maintaining logic across these different instances of the IR is a challenge for evaluating near-term noisy QPUs.
\section{XACC}
We have put forth a new hybrid classical-quantum programming model, XACC (eXtreme-scale ACCelerator), that attempts to address many of the challenges discussed above. XACC has been specifically designed for enabling near-term quantum acceleration within existing classical computing applications and workflows. This model, and associated open-source reference implementation, follows the traditional classical co-processor model, akin to OpenCL or CUDA for GPUs, but takes into account the subtleties and complexities inherent to the interplay between classical and quantum hardware. XACC provides a high-level API that enables classical applications to offload work (represented as quantum kernels) to an attached quantum accelerator in a manner that is agnostic to the quantum programming language and the quantum hardware. This enables one to write quantum code once, and perform benchmarking, verification and validation, and performance studies for a set of virtual (simulators) or physical quantum hardware. To achieve interoperability and portability, XACC defines four primary abstractions or concepts: quantum kernels, intermediate representation, compilers, and accelerators. Quantum kernels are C-like functions that contain code intended for execution on the QPU. These kernels are compiled to the XACC intermediate representation (IR), an object model that is key for promoting the integration of a diverse set of languages and hardware. The IR provides four main forms for use by algorithm programmers: (1) an in-memory representation and API, (2) an on-disk persisted representation, (3) human-readable quantum assembly representation, and (4) a control flow graph or quantum circuit representation. On top of this IR infrastructure, XACC defines further extension points for tools that pre-process the IR instance, post-process the IR instance, and a post-processor workflow layer that adjusts or corrects QPU results. The IR is produced by realizations of the XACC compiler concept, which delegates to the kernel language’s appropriate parser, compiler, and optimizer. Finally, XACC IR instances (and therefore programmed kernels) are executed by realizations of the Accelerator concept, which defines an interface for injecting physical or virtual quantum accelerators. Accelerators take this IR as input and delegate execution to vendor-supplied APIs for the QPU (or API for a simulator). The orchestration of these concepts enable an expressive API for quantum acceleration of classical applications. 

Currently, XACC provides support for a number of low-level quantum programming languages and both physical and virtual (simulators) hardware instances. XACC supports kernels written in Scaffold, Quil, as well as high-level domain-specific languages encoding second-quantized fermionic Hamiltonians. XACC provides a unified access to the Rigetti Forest infrastructure (simulators and the 19 qubit QPU) and the IBM Quantum Experience (simulators and 5 and 16 qubit QPUs). To demonstrate its use across computing models, the XACC interfaces are general enough that XACC also supports the D-Wave quantum annealer.

The overall goal of XACC is to provide an integration platform that enables researchers, engineers, and students to leverage near-term quantum computing architectures as part of existing classical computational workflows in a manner that fits the work at hand. Users should be able to express programs in a high-level domain-specific language that fits the science application and targets any available quantum computer. In this way, XACC can enable quantum computer science educational efforts, and for researchers, can enable benchmarking, verification, and profiling across a wide variety of QPU types.

\subsection{XACC and Portability}
Overall code portability is at the heart of the XACC design. XACC is the first platform to provide a robust and polymorphic intermediate representation object model that enables programming across quantum computing models (gate or adiabatic). This IR allows XACC to be language-agnostic, such that a code written in one language can be compiled to an IR instance and then mapped to a representation amenable for execution on a completely different QPU. This mapping step can be accomplished with user-specified IR transformation implementations. For example, imagine one writes quantum code for one architecture with a given qubit connectivity and wants to run it on another architecture with a different connectivity structure. XACC handles the execution of appropriate transformations on the IR that insert swap gate instructions to insure that two qubit interactions can be executed on the new architecture. This general IR transformation infrastructure makes sure that true code portability can be achieved in the case of hybrid classical-quantum computing. It provides a mechanism for future benchmarking of quantum computers via a number of different application types.

\subsection{XACC and Error Mitigation}
The XACC interfaces can also be leveraged to provide certain error mitigation strategies. Most novel error mitigation schemes involve some sort of classical pre-processing of the quantum program, followed by execution and post-processing of the result based on the pre-processing action. This type of workflow fits in very well with the XACC IR infrastructure. Classical pre-processing can be achieved through an implementation of the IR pre-processing interface. This interface takes an IR instance in, pre-processes or otherwise modifies this IR in some way, and then outputs a functional instance of some post-processing step that is stored by the XACC framework and executed after QPU execution. This is an ideal setup for mitigating the QPU qubit readout errors discussed in Section \ref{sec:noise}. XACC provides a pre-processor implementation that prepends the IR instance with extra quantum kernels that measure the overall probability of a bit flip error on a given qubit. Those results are stored and used by a post-processor functional instance, which is applied to the qubit measurement results after execution. This post-processor uses the bit flip probabilities to shift and scale resultant observable expectation values to more accurate values.

This workflow could be applied to other forms of error mitigation, requiring simple interface implementations that pre-process quantum IR and post-process QPU qubit measurement results.

\subsection{XACC and Access Models}
The XACC platform enables multiple forms of user access via the host language the framework is written in and its overall platform, or system context, model. XACC is written in C++ - a foundational language that enables bindings to many other programming languages. This greatly facilitates a user access to available QPUs since users are not tied to a language they are not familiar with. As of this writing, XACC provides Python bindings and has planned support for Fortran due to its wide adoption in classical high-performance scientific computing.

XACC implements a client-server access model designed to support both remote and local user access models, as discussed in Section \ref{sec:client}. Implementations of the XACC accelerator construct are currently available for QPUs with access to a REST client that enables HTTP POST/GET operations to affect execution of quantum programs on the remotely hosted service. The XACC client-server model also enables access to a local access model by using service invocations that can be routed to the local host computing system. XACC is currently ready to enable local QPU-access models simply by redirecting the accelerator URL to the locally hosted QPU driver server.


\subsection{XACC Compilation Workflow}
\label{sec:xworkflow}
The XACC compilation process directly implements the workflow layers described in Fig.~\ref{fig:workflow}. The XACC compiler interface takes quantum kernels as input and outputs a representative IR instance. This IR instance is then passed through any requested IR Preprocessors - which can be implementations for readout-error mitigation of the IR program, or mapping logical qubits to more appropriate physical qubits. This layer is extensible for future preprocessing implementation steps. The XACC IR Preprocessor takes as input the IR instances but produces a new post-processing function instance that is queued up for execution after the primary QPU execution. 

Next, the IR is passed to any requested optimization routines. An example implementation for this is a mechanism for simplifying the program to use fewer resources on the hardware. Next, transformations are executed that make the program amenable for execution on the hardware (to ensure requested two-qubit gates are available in hardware, and, if not, apply swap gate transformations, for example). Finally, the IR is mapped to the appropriate low-level machine instruction set implemented in hardware.

This compiled result is sent off for execution on the QPU, and the resultant bit strings are brought back and passed through any post-processing functional instances produced by the preprocessing layer. This can be, for example, a post-processor that updates observable expectation values based on error probabilities added to the overall computation via a previous IR Preprocessor.

\section{Application Examples}
In this section, we review a recent example application programmed and executed using XACC that was hardware agnostic and implemented the various layers of the quantum compilation workflow. This example made use of the variational quantum eigensolver (VQE) algorithm, which relies on the variational principles of quantum mechanics to find the minimal energy quantum state under a given Hamiltonian. The algorithm is relatively simple and has the advantage of permitting even short-depth circuits to address interesting application scenarios. Here we provide a brief overview of this application and detail how XACC addresses the various challenges detailed in this work.

\subsection{Nuclear Binding Energy}
We recently undertook the computation of the binding energy of the deuteron via cloud quantum computing resources using the variational quantum eigensolver hybrid algorithm \cite{deuteronarxiv}. Using XACC, we were able to program the problem in a hardware-agnostic manner and target available superconducting circuit gate model quantum computers from IBM and Rigetti \cite{noauthor_ibm_nodate,rigettiqaoa}. Via the compiler workflow discussed in Sections \ref{sec:workflow},\ref{sec:xworkflow}, we were also able to provide minimal error mitigation that corrected for qubit measurement readout errors. This work represented the first variational quantum eigensolver computation done through a remote access model, and highlighted the need for future local access models that take advantage of application job queues instead of individual QPU execution queues. This remote access model severely hindered the work and demonstrates the need for vendors to research and implement local access models that enhance or enable variational scientific quantum simulation. 
\par
We leave the low-level technical details of the deuteron work to \cite{deuteronarxiv}, but at a high level, we leveraged a pionless effective field theory in a discrete variable representation using the familiar harmonic oscillator basis. We considered cutoffs of that basis at $N=2,3$. Here we discuss the $N=2$ case for brevity, which employed the following Hamiltonian
\begin{align}
H_2 &= 5.906709 I +0.218291Z_0 -6.125 Z_1 \nonumber\\
& -2.143304 \left(X_0 X_1 + Y_0Y_1\right).\label{H2}
\end{align}
Dictated by this Hamiltonian, we performed measurements of our QPU after application of a unitary coupled cluster circuit composed of a single variational parameter $\theta$. Computing the ground state energy required looping over various $\theta$ parameters as part of a classical non-linear optimization scheme until convergence criteria were met. 
\par
\begin{tcolorbox}
\begin{lstlisting}[style=tcblatex,caption={XACC Kernels for Deuteron VQE},captionpos=t]
__qpu__ ansatz(AcceleratorBuffer b, double t0) {
    X 0
    RY(t0) 1
    CNOT 1 0
}
__qpu__ z0(AcceleratorBuffer b, double t0) {
    ansatz(b,t0)
    MEASURE 0 [0]
}
__qpu__ z1(AcceleratorBuffer b, double t0) {
    ansatz(b,t0)
    MEASURE 1 [1]
}
\end{lstlisting}
\begin{lstlisting}[style=tcblatex]
 __qpu__ x0x1(AcceleratorBuffer b, double t0) {
    ansatz(b,t0)
    H 0
    H 1
    MEASURE 0 [0]
    MEASURE 1 [1]
}
__qpu__ y0y1(AcceleratorBuffer b, double t0) {
    ansatz(b,t0)
    RX(1.57079) 0
    RX(1.57079) 1
    MEASURE 0 [0]
    MEASURE 1 [1]
}
\end{lstlisting}
\label{lst:kernels1}
\end{tcolorbox}
\par
The code for this work was written in as XACC quantum kernels in the Quil language from Rigetti \cite{quil} and is shown in Listing 1. Through XACC, this code was immediately portable to IBM (as well as a number of simulators), thus overcoming the portability challenge for near-term quantum programming and computation. 
\par
\begin{figure}[ht]
\centering
\includegraphics[width=3.3in]{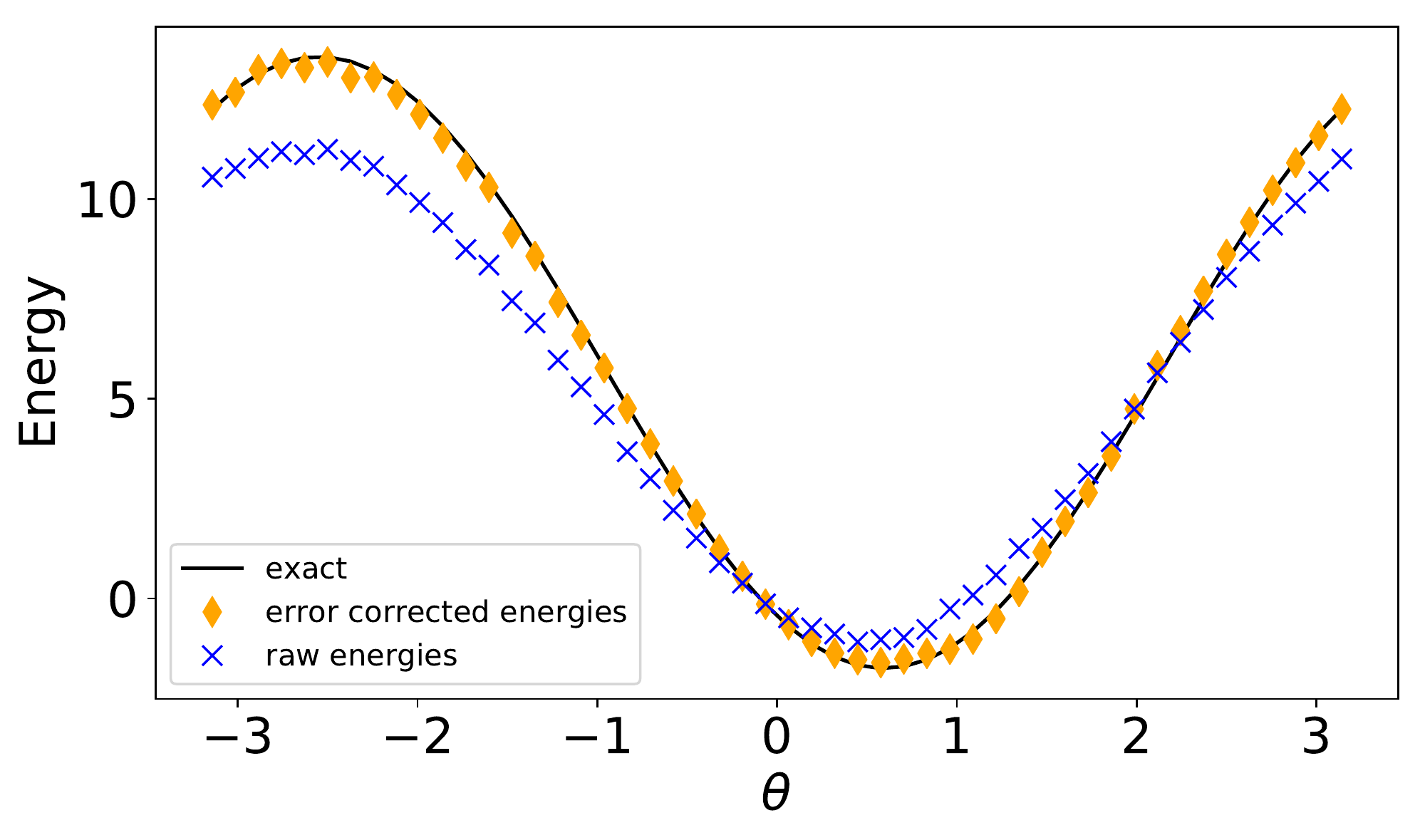}
\caption{Energy as a function of the variational parameter for the deuteron $N=2$ Hamiltonian with and without readout error correction.}
\label{fig:deuteron_error}
\end{figure}

One common source of error discussed in Section \ref{sec:noise} are due to systematic errors in reading out the state of an individual qubit. These types of errors were discussed in the supplemental information of \cite{kandala2017}. For this work we automated this error mitigation strategy as part of the XACC compiler workflow. We implement the XACC IR Preprocessor extension interface to append measurements of each qubit that provide probabilities that the qubit was in a state of 0 when a 1 was expected, and vice versa. This IR Preprocessor implementation then returns a post processing function instance that is executed after QPU execution that leverages these probabilities to shift and scale observable expectation values. The plot in Figure \ref{fig:deuteron_error} shows the energy as a function of the variational parameter, with and without this readout error mitigation preprocessor execution. Clearly, automating this sort of error mitigation will provide more reliable results with minimal costs to those adopting quantum computing as part of their scientific computing workflows.
\section{Conclusion}
Hybrid computing systems offer novel platforms to integrate emerging QPU with conventional programming methods. However, there are several challenges that arise from these noisy devices whose performance not yet well understood. In this contribution, we have outlined many of the technical issues faced by quantum program developers adopting to client-server model for remote access across multiple technologies and vendors. In addition to new needs for device-level information, current programmers also face obstacles in code portability, tool integration, program validation, and workflow development.
\par
In the context of these challenges, we have described how the XACC programming framework provides new methods for inter-operable program and tool development as well as support for new access client-server access models based on local QPU systems. The framework itself is hardware agnostic and, therefore, meant to provide a generalized approach to quantum programming. This contrast with the diversity of vendor-specific stacks and domain-specific languages underdevelopment. We anticipate that both efforts, specialized and generalized, are needed to ensure strong and robust growth of the quantum computing ecosystem. The ongoing co-design of hardware and software to mitigate gate noise and execution errors will continue to require close coordination.
\section*{Acknowledgment}
This work has been supported by the Laboratory Directed Research and Development Program of Oak Ridge National Laboratory and the US Department of Energy (DOE) Office of Science Advanced Scientific Computing Research Early Career Research Award. 

\ifCLASSOPTIONcaptionsoff
  \newpage
\fi



\bibliographystyle{IEEEtran}
\bibliography{icrc2018}
\end{document}